# Self-Driven Highly Responsive PN Junction InSe Heterostructure Near-Infrared Light Detector

CHANDRAMAN PATIL,[1] CHAOBO DONG,[1] HAO WANG,[1] BEHROUZ MOVAHHED NOURI,[1,2] SERGIY KRYLYUK,[3] HUAIRUO ZHANG,[3,4] ALBERT V. DAVYDOV,[3] HAMED DALIR,[1,2] AND VOLKER J. SORGER,[1,2,*]

[1]Department of Electrical and Computer Engineering, George Washington University, Washington, DC 20052, USA

[2]Optelligence LLC, 10703 Marlboro Pike, Upper Marlboro, MD 20772, USA

[3]Materials Science and Engineering Division, National Institute of Standards and Technology Gaithersburg, MD 20899, USA

[4] Theiss Research, Inc., La Jolla, CA 92037, USA

*sorger@gwu.edu

**Abstract:** Photodetectors converting light signals into detectable photocurrents are ubiquitously in use today. To improve the compactness and performance of next-generation devices and systems, low dimensional materials provide rich physics to engineering the light-matter interaction. Photodetectors based on two-dimensional (2D) material van der Waals heterostructures have shown high responsivity and compact integration capability, mainly in the visible range due to their intrinsic bandgap. The spectral region of near-infrared (NIR) is technologically important featuring many data communication and sensing applications. While some initial NIR 2D material-based detectors have emerged, demonstrations of doping-junction-based 2D material photodetectors with the capability to harness the charge-separation photovoltaic effect are yet outstanding. Here, we demonstrate a 2D p–n van der Waals heterojunction photodetector constructed by vertically stacking p-type and n-type indium selenide (InSe) flakes. This heterojunction charge-separation-based photodetector shows a three-fold enhancement in responsivity in the near-infrared (NIR) spectral region (980 nm) as compared to photoconductor detectors based on p- or n- only doped InSe. We show that this junction device exhibits self-powered photodetection operation, few pA-low dark currents, and is about 3-4 orders of magnitude more efficient than the state-of-the-art foundry-based devices. Such capability opens doors for low noise and low photon flux photodetectors that do not rely on external gain. We further demonstrate millisecond response rates in this sensitive zero-bias voltage regime. Such sensitive photodetection capability in the technologically relevant NIR wavelength region at low form factors holds promise for several applications including wearable biosensors, 3D sensing, and remote gas sensing.

1.  **Introduction**

2D semiconducting materials due to their bandgap have been studied as promising photodetector materials, by changing the layer numbers or forming van der Waals (vdW) heterostructures, owing to their high responsivity, fast response time, broadband photodetection, photo-detectivity, and low dark current noise [1]-[7]. The operation of these high-performance devices demands high bias voltage leading to large power consumption due to the Schottky barrier potential and poor photogenerated carrier collection. This limits technological applications for remote operation conditions under extreme environments, biomedical sensing, and portable devices [8]-[11]. However, self-driven photodetectors are promising devices to solve energy consumption issues where the photon energy is higher than the bandgap of the material for a better signal-to-noise ratio. Recently, few self-driven 2D material-based photodetectors have been demonstrated in [9], [12]-[14]. III-VI group 2D materials (InSe, GaSe, and GaTe) have been recently studied for light-matter interaction properties for optoelectronic applications [15]-[17]. 2D InSe, with its direct bandgap (~1.25 eV) [18] has recently been investigated showing higher ultrasensitive photodetection characteristics than other 2D semiconducting materials such as $MoS_2$ and $WSe_2$ [19]-[22]. For p- and n- doped materials used in heterostructure devices form atomically sharp p-n junctions [23]-[25].

    A thorough study has been performed on InSe-based photodetectors in the visible spectrum but is yet to be explored in detail for near-infrared (NIR) applications [26], [27]. Optoelectronic devices for detection, modulation, and sensing applications in the mid-infrared spectrum are also limited by the lack of on-chip sources and the high cost of production [28]-[30]. The optoelectronic devices based on InSe are usually designed in transistor configuration that requires high electrical bias gating and bias voltage [19], [27], [31]. Furthermore, the NIR wavelength at 980 nm is widely used for optical humidity sensing fields such as indoor air quality, industrial production process control, and agricultural instrumentation [32], [33]. Also,

with the technological advancement in building autonomous vehicles, LiDAR-based devices usually are designed at 940 – 980 nm wavelengths for short to medium range positioning and mapping [34], [35]. Due to a direct bandgap (1.25 eV) [18], InSe is an attractive material for manufacturing optoelectronic devices in this range.

Here, we demonstrate a self-driven p-InSe/n-InSe heterostructure photodetector for NIR applications. The p- and n- doped 2D InSe flakes form the vertical heterostructure stack showing a ~3 times improvement in responsivity and ~3.5 times lower response time as compared to p- or n-type InSe photodetectors thus demonstrating a novel fast and sensitive InSe heterojunction-based NIR photodetector suitable for low power optical sensors or detector devices.

## 2. Results and Discussion

Enhancing light-matter interaction by building heterostructures using 2D materials paves the way toward building high-performance and energy-efficient compact photodetectors [36]. The 2D material-based photodetectors are often limited by the high bias voltage required for photon-generated carrier collection due to high resistance and the Schottky barrier [37]. Here, we demonstrate a pn junction-based photodetector using InSe for NIR detection or sensing applications as seen in **Fig. 1a**. The 2D InSe flakes are mechanically exfoliated from bulk crystals grown by the vertical Bridgman method and transferred precisely on prefabricated Au/Ti metal contacts using the novel 2D material transfer system discussed in [38]. The optical microscope image of the device is shown in **Fig. 1b**. A cross-sectional scanning transmission electron microscopy (STEM) image of the photodetector is shown in **Fig. S4** of the supporting information. The pn InSe heterojunction is formed by the physical contact between Sn-doped InSe (n-type) and Zn-doped InSe (p-type) materials. The band diagram for the device structure can be seen in Fig. 1c where the built-in heterojunction potential helps in collecting the photogenerated carriers in the absence of external electrical bias potential. The material quality was assured after transfer using Raman spectroscopy by monitoring the relative intensities of Raman active modes at $A^1_{1g}$, $A^2_{1g}$, and $E^1_{2g}$ [18].

The photovoltaic properties of the p-InSe/n-InSe heterojunction, p-InSe, and n-InSe based photodetector device is studied at 980 nm under vertical illumination using a free-space optical setup. **Fig. 2a** shows the current-voltage (I-V) characteristics for photocurrent measurement of the pn-InSe junction, p-InSe, and n-InSe, where the illumination power is 111.6 µW. The pn-junction device exhibits an order of magnitude higher photocurrent as compared to the p- and n- InSe devices, thus indicating a higher photo-absorption at 980 nm for the vdW heterojunction characteristics. **Fig. 2b** shows the power-dependent photocurrent response of the pn junction device. It can be observed that the photocurrent saturates after reaching high optical incident power intensity. The photovoltaic effect is attributed to the built-in electric field in the heterojunction depletion region. **Fig. 2c** shows the dark current mapped at different voltages for the device. This shows a very small change in the dark current with an increase in the bias voltage leading to low electrical energy loss. As a result of the built-in potential of the junction, the device can be operated under no external bias for collecting photo-generated carriers. **Fig. 2d** shows the relationship between the photocurrent and the input optical power for pn-InSe heterojunction, p-InSe, and n-InSe devices under zero bias voltage at 980 nm light illumination. It can be seen that the built-in potential in the pn junction device shows higher photocurrent generation as compared to the non-junction (p- and n-) type devices.

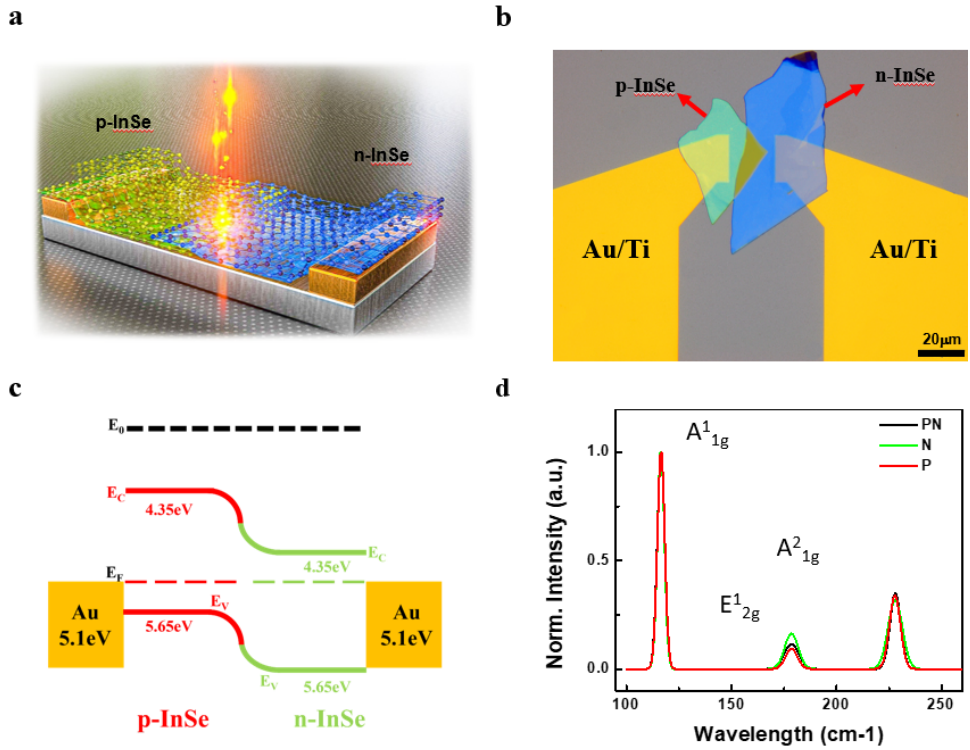

**Fig. 1. PN heterojunction n-InSe/p-InSe photodetector** (a) A schematic representation of the n-InSe/p-InSe van der Waals p-n junction photodetector. (b) An optical micrograph image of the device (top view), where n-InSe was stacked on the top of the p-InSe flake, transferred on Au/Ti electrodes using the 2D printer technique. (c) Band diagram for p-InSe (red), n-InSe (green) and Au contact (yellow). (d) The Raman spectra of the p-InSe, n-InSe, and the junction regions. All material-associated Raman peaks of p-InSe, n-InSe, and pn InSe junction are observed to show peak positions and relative intensity associated with out-of-plane vibrational modes ($A^1_{1g}$ and $A^2_{1g}$) and in-plane vibrational modes ($E^1_{2g}$).

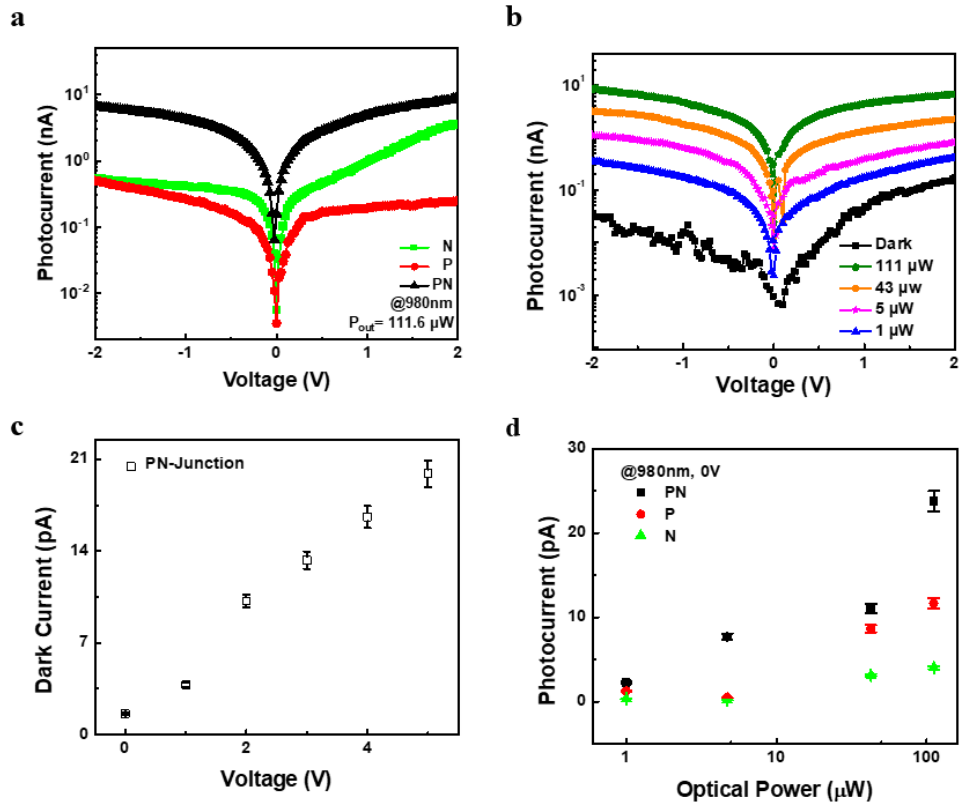

**Fig. 2. The photovoltaic characteristics of pn-InSe heterojunction** (a) Typical I–V characteristics of p (red), n (green), and pn junction (black) 2D InSe photodetector. (b) The I-V characteristics of the pn-InSe heterojunction under different optical input power show saturation of photocurrent at higher optical power. (c) Dark current mapping under bias voltage for pn junction device indicates picoamps range current. This exhibits a noise equivalent power (NEP) of ~2nW/Hz$^{0.5}$ at zero bias. (d) The corresponding fitting curves for the relationship between the photocurrents and the optical power of the p-InSe/n-InSe heterojunction in zero biased voltage, in 980 nm light condition.

Further, the photodetector devices were tested for responsivity as a function of wavelength from 800 nm to 900 nm (supercontinuum source) and at 980 nm (diode laser) for NIR photodetection. **Fig. 3a** shows the responsivity ($R_\lambda=[I_{ph}-I_{dark}]/P_{in}$) of pn heterojunction, p-, and n- InSe at the zero-bias voltage for pn device and at 2 V for p- and n- devices. The pn junction device shows enhanced responsivity along the NIR wavelength as compared to n- and p- devices by 3.03 times. It can be found that the maximum responsivity of p-InSe/n-InSe heterojunction is about 0.5 mAW$^{-1}$ at 980 nm under zero bias. The photoluminescence (PL) spectra of the p-InSe, n-InSe, and pn-InSe heterojunction from 800 nm to 1000 nm were recorded using a 532 nm laser for excitation. The exciton recombination peak at 980 nm (1.265 eV) in the PL spectrum shows the intensity of pn-InSe heterojunction higher than that of p-InSe and n-InSe by ~5.9 times at 980 nm as seen in **Fig. 3b**. It is known that the exciton PL is very sensitive to the presence of defects and surface contaminants. The enhanced PL intensity at the pn junction, which is due to the increased photoexcitation volume, provides evidence of a clean interface between the p- and n-InSe flakes. The Raman spectroscopy mapping of the pn junction stack also confirmed the clean heterojunction formed between the two flakes (**Fig. 3c**). No additional peaks were observed due to contaminants or defects produced in the material during device fabrication [39].

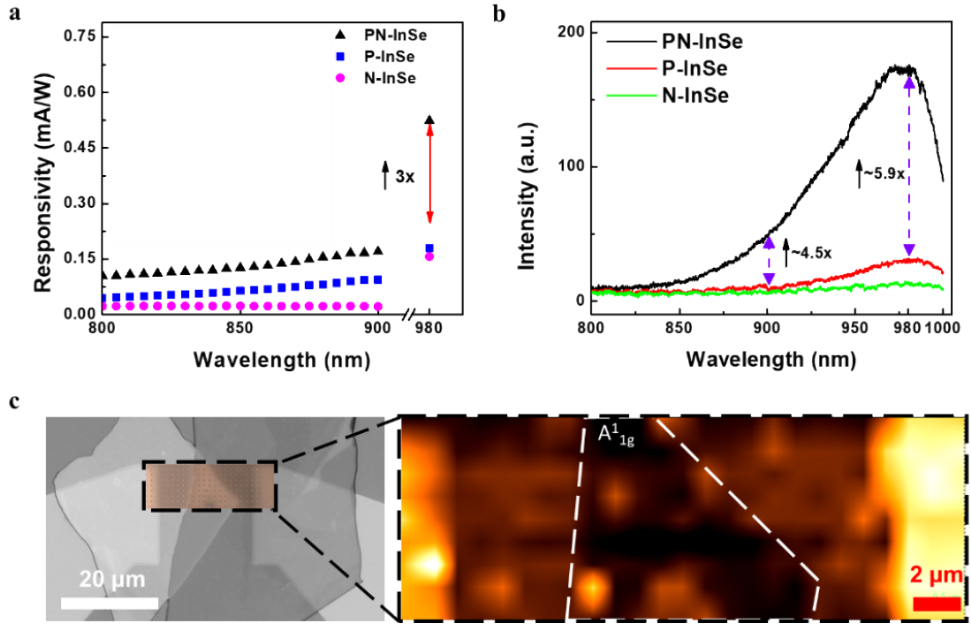

**Fig. 3. Spectral characterization of pn-InSe heterojunction** (a) Experimental spectra of responsivity of p-InSe, n-InSe, and pn-InSe heterojunction at 0 V under 800–900 nm and 980 nm light illumination. (b) Photoluminescence spectra of p-InSe, n-InSe, and p-InSe/n-InSe heterojunction show strong agreement with the responsivity spectra. An enhancement of 4.5 times and 5.9 times in intensity is observed at 900 nm and 980 nm, respectively. (c) The quality of the heterojunction created was assessed at the $A^1_{1g}$ Raman peak at the physical position of the heterojunction region (white dashed region), as shown by Raman mapping analysis. A 532 nm laser was used for excitation.

The photoresponse of the pn-InSe photodetector is expected to be higher than the p- and n-InSe due to the built-in potential inside the heterojunction. The on/off switching response for pn-InSe (0 V), p- (-2 V), and n- (2 V) photodetector under 980 nm (111.6 µW) illumination was measured by modulating the optical source power supply. With an increase in the external electric field, the charge density in the InSe is high leading to a decrease in mobility of photo-generated carriers because of the decrease in carrier drift velocity. The rise and fall time constants of the pn junction device show a faster response as compared to the p- and n-only InSe photodetectors **(Fig.4)**. Due to the absence of junction for p- and n- InSe devices, the photocarrier collection is possible only after applying an electrical bias. Therefore, both the devices were tested at 2 V bias for the response time measurement, unlike the built-in electric field of the p-n junction, light-excited electron-hole pairs can be more effectively separated due to the photovoltaic effect versus relying on photoconductivity for the n- and p-only junction devices. The response time of the pn-InSe device shows a rise time/fall time ($\tau_r/\tau_f$) response of 8.3 ms/9.6 ms **(Fig.4a)**, which is reduced as compared to n- and p- InSe device by 18 ms/40 ms **(Fig.4b)** and 22 ms/29 ms **(Fig.4c)**, respectively. This is an interesting result since it shows the engineerability for designing fast response photodetectors by stacking the 2D materials and enhancing the optoelectronic properties of the material for various applications.

The response time of this device is limited by trap states generated in-between material to the substrate and 2D-2D interlayer boundaries. Such trap states help achieve signal detection effectively due to the accumulation of charges but adversely affect the speed response of the device. The trap states generated between the substrate and 2D material majorly contribute to the slow response. This can be improved further by building the device on a 2D substrate material like hBN for preserving the mobility of the material and achieving a faster response time. The low dark current of the device provides a stable noise floor owing to the presence of built-in potential. As seen in **Fig. 2c**, the minute change in dark current with an increase in voltage shows that the device can be operated over a varying range of bias voltage without increasing the noise floor of the device. Such device engineering would provide better solutions to lower power sensing and wearable electronics applications.

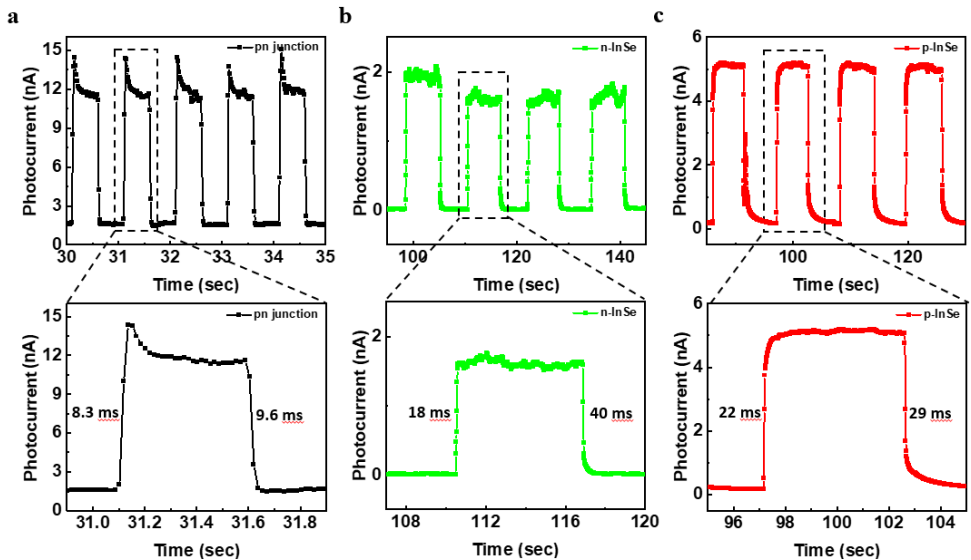

**Fig. 4. Time dependent photoresponse of** (a) pn-InSe photodetector at 0 V with rise time/fall time ($\tau_r/\tau_f$) of 8.3 ms/9.6 ms., (b) n-InSe at 2 V with $\tau_r/\tau_f$ of 18 ms/ 40ms, and (c) p-InSe at -2 V with $\tau_r/\tau_f$ of 22 ms/29 ms.. Under 980 nm light illumination.

Lastly, the pn-junction device was tested for photoresponse and photocurrent mapping at external bias voltages. The gradient map demonstrates a faster response of the device at zero bias and at sub-volt forward bias voltages (**Fig. 5a**). The nature of the shape of the gradient is governed by the usually observed pn diode I-V characteristics. The separation of photo-generated electron-hole pairs is better under the influence of external bias voltage. However, the device suffers from high charge density leading to lower drift current velocity in the material. Therefore, the time response and sensitivity of the device are adversely affected by applied external bias. This device shows promising performance for sensing applications in the NIR spectral region owing to its extremely low dark current and response rate at zero external bias. A summary of the tradeoff between the response time of the device and operational voltages is represented in **Fig. 5b** based on the current and previously reported photodetector devices for NIR applications [40]-[48]. Additional comparison of performance metrics is shown in Table T1 from previously published work.

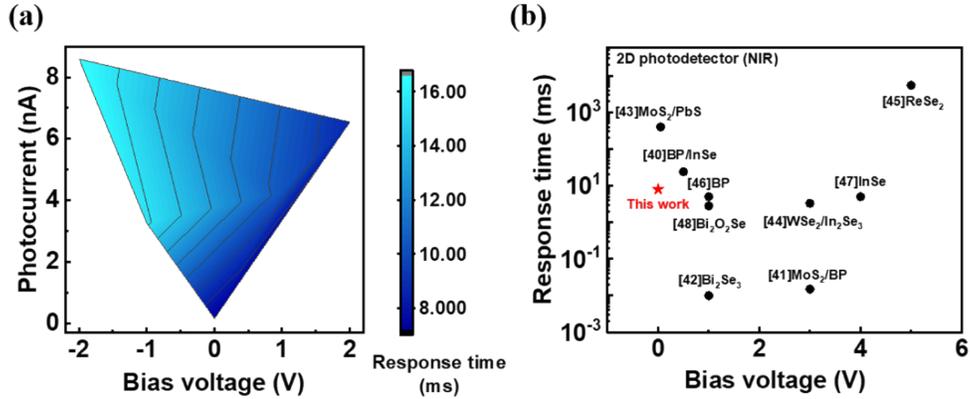

**Fig. 5. Photoresponse mapping under external bias** (a) Photoresponse at different bias voltage (-2 V to 2 V) with photocurrent of pn-InSe under 980 nm illumination demonstrated the fastest response under 0 V bias condition. (b) Comparison of the photoresponse under the external bias of 2D photodetectors for NIR applications discussed in the literature.

| Materials | λ (nm) | Responsivity (mAW$^{-1}$) | Bias voltage (V) | $I_{dark}$ (nA) | $\tau_r/\tau_f$ (ms/ms) | Ref. |
|---|---|---|---|---|---|---|
| InSe | 700 | $3.06 \times 10^6$ | 5 | ~ 200 | 5/8 | [47] |
| Doped-InSe | 980 | $7.87 \times 10^6$ | 1 | 800 | 0.45/6× 10$^3$ | [27] |
| Bi$_2$Se$_3$ | 1456 | 274 | 1 | ~ 600 | 54/47 | [42] |
| BP/InSe | 633 | $10^7$ | 0 | ~100 | 24/32 | [40] |
| Gr/GaAs | 980 | 5.97 | 0 | 0.275 | ~40 | [13] |
| Te/Ge | 980 | 522 | 0 | -50 | 14/0.105 | [49] |
| Ge bulk | 980 | 2214 | -0.1 | ~5 | - | [50] |
| MoS$_2$ | 473−2712 | 47.5 | 1 | ~400×10$^3$ | 10/16 | [51] |
| InGaAs | 1100-2000 | 3.5× 10$^6$ | 0.5 | 144 | 70/280 | [52] |
| **pn-InSe** | 980 | 0.5 | 0 | 1.5×10$^{-3}$ | 8.3/9.6 | This work |

Table T1. Comparison of the performance of our n-InSe/p-InSe photodetector with other NIR photodetectors

## 3. Conclusion

A self-driven photodetector was realized for NIR applications at 980 nm and an extremely low dark current in the range of a few picoamps was achieved. The photoresponse of the 2D InSe was enhanced by building a pn vdW heterojunction using p- and n- doped InSe showing an increase of ~3 times in responsivity as compared to the control n-InSe and p-InSe photodetector devices. The pn device also exhibits a faster response which is ~3.5 times lower than a p- or n-type InSe photodetector, thus demonstrating a novel fast, and sensitive InSe heterojunction-based NIR photodetector suitable for low power optical sensors or detector devices. Such performance of this device shows a high potential for realizing a NIR photodetector for sensing and optical applications. Furthermore, by engineering the interlayer stacking to match closely with the lattice structure and improving the metal contact Schottky barrier, it is expected that the device performance in the NIR region will be further improved. Further investigations are required to determine the chemical and mechanical material stability of the InSe heterojunction structure under various environmental effects for applications like remote sensing and biological sensing.

## 4. Methods

### 4.1 Crystal Growth and Characterization

InSe single crystals were grown by the vertical Bridgman-Stockbarger method from non-stoichiometric polycrystalline In$_{1.04}$Se$_{0.96}$ powders. Tin (2 at. %) and Zn (as ZnSe, 0.3 at.

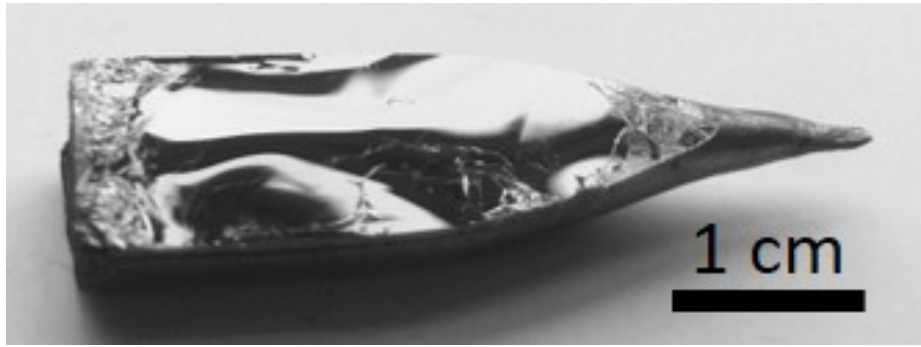

**Fig. 6.** Photograph of a p-InSe ingot cleaved parallel to the c-plane.

%) were added during the powder synthesis to obtain n- and p-typed doped crystals, respectively. The InSe melt in a graphitized quartz ampoule was equilibrated at 720 °C for several hours, then the ampoule was translated across a temperature gradient at a rate of 0.5 mm/h. A representative picture of an InSe ingot is shown in **Fig. 6**. The crystal structure was determined using annular dark-field scanning transmission electron microscopy (ADF-STEM) which confirmed the γ phase of InSe (**Fig. 7**). Electrical properties were determined from Hall effect measurements in the Van der Pauw geometry at room temperature. For n-InSe (p-InSe), the carrier concentration and Hall mobility are $1.5\times10^{16}$ cm$^{-3}$ and 616 cm$^2$V$^{-1}$s$^{-1}$ ($7.9\times10^{13}$ cm$^{-3}$ and 43 cm$^2$V$^{-1}$s$^{-1}$), respectively.

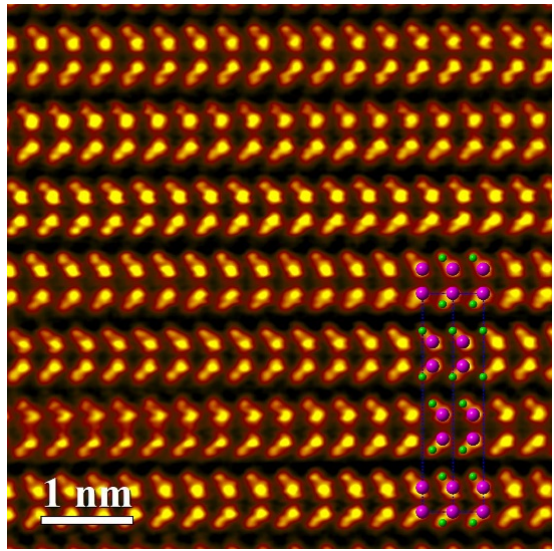

**Fig. 7.** Atomic resolution annular dark-field scanning transmission electron microscopy (ADF-STEM) image of InSe showing a good match with the overlapped atomic model of γ-InSe. Green and purple dots in the atomic model represent Se and In atoms, respectively.

*4.2 Device Fabrication*

The InSe pn junction heterostructure is formed using 2D flakes exfoliated from the bulk crystals and transferred using the micro stamp-assisted transfer system on prefabricated electrical contacts [38]. Here, the p- and n-type InSe flakes were obtained by mechanical exfoliation from the bulk crystals using Nitto tape and were transferred to thin (17 mils) PDMS film. Briefly, this method comprises a micro stamper to transfer the material from PDMS which can be aligned under a microscope precisely at any desired device location via micro-positioners as seen in **Fig. 8**. We scan the PDMS film from the top to find a flake of interest under the PDMS film and transfer it at a targeted location (between metal contacts) without any cross-contamination of other 2D material flakes on PDMS on the substrate. The micro-stamper guides a flake to the target location and transfers it onto the substrate by pressing the PDMS film gently. The sample is rinsed cleaned using acetone and iso-propyl alcohol and dried using nitrogen gas after each transfer of layer. The sample is soft-baked at 85 °C for 2 min (APR). We added a video link in the manuscript as a reference, with the published 2D printer work [53], [54] which describes our rapid, clean, and precise transfer method. Note, the video shows rapid integration of van der Waals heterostructure on top of a Si photonic chip (IMEC) by precisely placing two different 2D materials (2H-MoTe$_2$ and hBN), by using our novel 2D printer technique for on-chip photonic devices. The electrical contacts were formed using electron beam lithography for a channel length of 20 μm. The metal Ti/Au (5 nm/45 nm) contacts were formed by deposition using the electron beam evaporation method. The lift-off was performed using acetone at room temperature followed by rinsing in isopropyl alcohol and nitrogen drying.

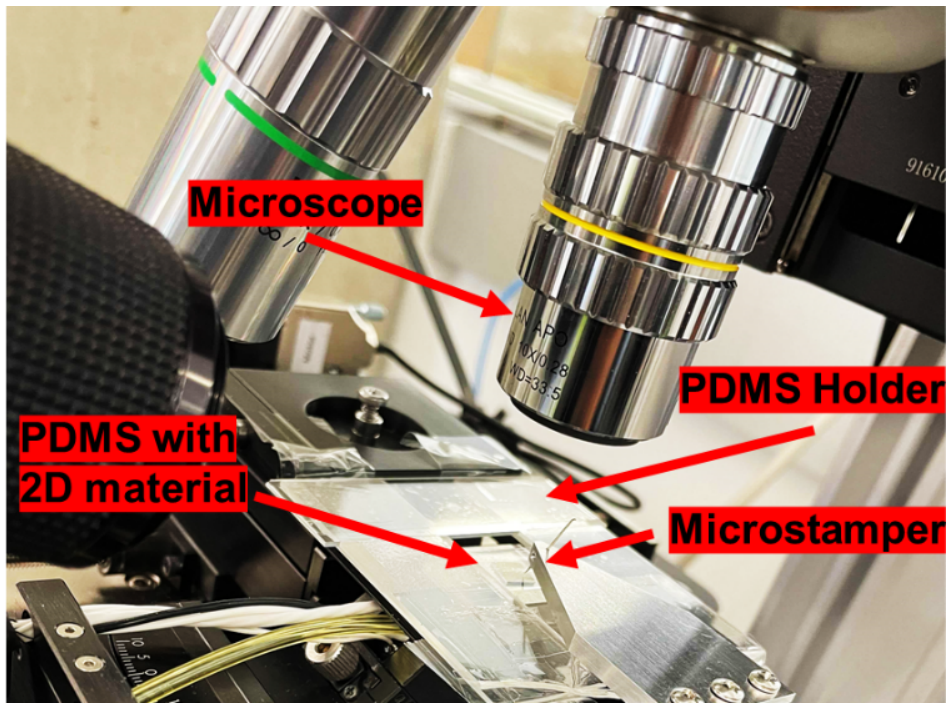

**Fig. 8.** 2D Printer transfer setup used for building the pn heterojunction photodetector

*4.3 Scanning transmission electron microscopy (STEM) characterization*

A FEI Nova NanoLab 600 dual-beam system (SEM/FIB) was employed to prepare an electron transparent cross-sectional sample. Electron-beam-induced deposition of 100 nm thick Pt was initially deposited on top of the 2D pn junction device to protect the sample surface, then followed by 2 μm ion-beam induced Pt deposition. To reduce Ga-ions damage, in the final stage of FIB preparation the sample was thinned with 2 kV Ga-ions using a low beam current of 29 pA and a small incident angle of 3 degrees. An FEI Titan 80-300 STEM/TEM equipped with a probe spherical-aberration corrector was employed to obtain annular dark field (ADF) STEM images with an operating voltage of 300 kV, probe convergence semi-angle of 14 mrad, and collection angle of 34-195 mrad. A typical cross-sectional ADF-STEM image of the pn InSe photodetector shown in **Fig. 9** suggests the 2D flakes have good contact with the Au/Ti electrode, as well as in the p-n junction region.

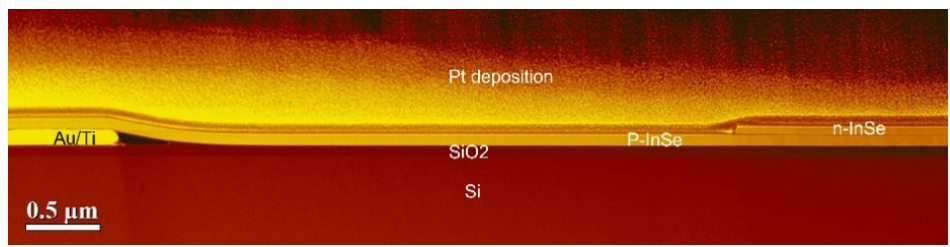

**Fig. 9.** Cross-sectional ADF-STEM image of the pn-InSe photodetector fabricated using the 2D printer transfer setup.

### 4.4 Device Experimentation

The experimental setup for measuring the 2D InSe pn junction heterostructure devices comprises a tunable (NKT SUPERCONTINUUM Compact) source and a fixed wavelength laser diode module (CPS980 Thorlabs, INC.) at 980 nm wavelength. The light beam was focused on the device using an objective lens. A source meter (Keithley 2600B) was used for electrical response measurements. The response time of the device was measured by modulating the electrical power supply to the laser. The Raman and photoluminescence measurements were performed at room temperature using a 532 nm laser. InSe crystal structure and interface quality of the pn photodetectors were examined using an FEI Titan 80-300 probe-corrected scanning transmission electron microscope (STEM) operating at 300 keV.

**Disclosures**

**Data availability.** Data underlying the results presented in this paper are not publicly available at this time but may be obtained from the authors upon reasonable request.

**Funding.** V.S. is supported by the AFOSR PECASE award (FA9550-20-1-0193).

**Acknowledgments.** A. V. D. and S. K. acknowledge support through the Material Genome Initiative funding allocated to the National Institute of Standards and Technology. H.Z. acknowledges support from the U.S. Department of Commerce, National Institute of Standards and Technology under the financial assistance awards 70NANB19H138.

Disclaimer: Certain commercial equipment, instruments, or materials are identified in this paper to specify the experimental procedure adequately. Such identification is not intended to imply recommendation or endorsement by